\documentclass[aps,pre,twocolumn,superscriptaddress,longbibliography]{revtex4-1}
\usepackage{multirow}
\usepackage{epsfig}
\usepackage{xcolor}
\usepackage{graphicx}
\usepackage{dcolumn}
\usepackage{bm}
\usepackage{amsmath}

\begin{document}

\title{A statistical inference approach to structural reconstruction of complex networks from binary time series}

\author{Chuang Ma}
\affiliation{School of Mathematical Science, Anhui University, Hefei 230601, China}

\author{Han-Shuang Chen}
\affiliation{School of Physics and Material Science, Anhui University, Hefei 230601, China}

\author{Ying-Cheng Lai}
\affiliation{School of Electrical, Computer and Energy Engineering, Arizona State University, Tempe, Arizona 85287, USA}

\author{Hai-Feng Zhang} \email{haifengzhang1978@gmail.com}
\affiliation{School of Mathematical Science, Anhui University, Hefei 230601, China}
\affiliation{Center of Information Support \&Assurance Technology, Anhui University, Hefei  230601,  China
}
\affiliation{Department of Communication Engineering, North University of China, Taiyuan, Shan'xi 030051, China}

\date{\today}

\begin{abstract}

Complex networks hosting binary-state dynamics arise in a variety of contexts. In spite of previous works, to fully reconstruct the network structure from observed binary data remains to be challenging. We articulate a statistical inference based approach to this problem. In particular, exploiting the expectation-maximization (EM) algorithm, we develop a method to ascertain the neighbors of any node in the network based solely on binary data, thereby recovering the full topology of the network. A key ingredient of our method is the maximum likelihood estimation of the probabilities associated with actual or non-existent links, and we show that the EM algorithm can distinguish the two kinds of probability values without any ambiguity, insofar as the length of the available binary time series is reasonably long. Our method does not require any {\em a priori} knowledge of the detailed dynamical processes, is parameter free, and is capable of accurate reconstruction even in the presence of noise. We demonstrate the method using combinations of distinct types of binary dynamical processes and network topologies, and provide a physical understanding of the underlying reconstruction mechanism. Our statistical inference based reconstruction method contributes an additional piece to the rapidly expanding ``toolbox'' of data based reverse engineering of complex networked systems.

\end{abstract}
\maketitle

\section{Introduction} \label{sec:intro}

Data based reconstruction of complex networked systems has been an active area
of research in network science and engineering with applications in a wide
range of disciplines~\cite{GDA:2002,GAR:2002,GDLC:2003,PG:2003,BDLCNB:2004,ECCBA:2005,BMADB:2006,YRK:2006,BL:2007,Timme:2007,TYK:2007,NS:2008,Sontag:2008,CMN:2008,WCHLH:2009,DZMK:2009,RWLL:2010,CHR:2010,YSWG:2010,LP:2011,HKKN:2011,ST:2011,YP:2011,WLGY:2011,WYLKG:2011,WYLKH:2011,YLG:2012,PYS:2012,WRLL:2012,BHPS:2012,SBSG:2012,SNWL:2012,SWL:2012,HBPS:2013,ZXZXC:2013,CLL:2013,TimC:2014,SWL:2014,SLWD:2014,SWFDL:2014,CLL:2015,SWWL:2016,LSWGL:2017}.
A tacit assumption in most existing works is that continuous-valued nodal
time series, either in continuous or discrete time, are available so that
various statistical measures can be computed for identifying the underlying
network structure. This has led to a diverse array of reconstruction
methodologies~\cite{GDA:2002,GAR:2002,GDLC:2003,PG:2003,BDLCNB:2004,ECCBA:2005,BMADB:2006,YRK:2006,BL:2007,Timme:2007,TYK:2007,NS:2008,Sontag:2008,CMN:2008,WCHLH:2009,DZMK:2009,RWLL:2010,CHR:2010,YSWG:2010,LP:2011,HKKN:2011,ST:2011,YP:2011,WLGY:2011,WYLKG:2011,WYLKH:2011,YLG:2012,PYS:2012,WRLL:2012,BHPS:2012,SBSG:2012,SNWL:2012,SWL:2012,HBPS:2013,ZXZXC:2013,CLL:2013,TimC:2014,SWL:2014,SLWD:2014,SWFDL:2014,CLL:2015,SWWL:2016,LSWGL:2017,MZL:2007}.
For example, from time series the traditional Pearson correlation can be
calculated to reveal the complex structure of the brain functional and
neural networks~\cite{ECCBA:2005,BMADB:2006,FRM:2008,SMTM:2008}.
Bayesian estimation has also been used for reconstructing neural
networks~\cite{Friston:2002,PP:2009}.
Based on continuous time series and knowledge about the nodal dynamical
equations, a delayed feedback control scheme can be designed to reveal the
network structure based on the principle of
synchronization~\cite{YRK:2006,YP:2011}. For
stochastic and nonlinear network dynamics that generate noisy, continuous
time series, situations can arise where the network matrix is directly
proportional to the dynamical correlation matrix that can be calculated
straightforwardly from the time series, leading to a class of
computationally efficient reconstruction
methods~\cite{WCHLH:2009,RWLL:2010,WRLL:2012,CLL:2013,CLL:2015}. When
the data amount is small, i.e., when only short (continuous or discrete)
time series are available, the principle of compressive sensing~\cite{CS:book}
can be exploited to develop frameworks for a variety of reconstruction
tasks~\cite{WLGY:2011,WYLKG:2011,WYLKH:2011,SWL:2014,SLWD:2014,SWFDL:2014,HSWD:2015,SWWL:2016,LSWGL:2017}. (See Ref.~\cite{WLG:2016} for a recent review.)
We note that, in most previous works, the values of the measured time
series are continuous in a range, regardless of whether time is continuous
or discrete.

In real systems, there are network dynamical processes that
generate binary time series, i.e., the values of the time series assume
only two possible values, e.g., 0 or 1. For example, in disease spreading on
a social network, the state of each node (person) can be conveniently
characterized as susceptible or infected~\cite{PCVV:2015}, which equally
applies to virus propagation on computer networks. In certain class of neural
networks, each node (neuron) can be classified as active or
inactive~\cite{KRA:2010}. In classical evolutionary game dynamics such as
the prisoner dilemma game, each node can be in one of the two states:
cooperation or defection~\cite{SF:2007,DCHW:2009}. In a political network,
the state of each node (the opinion of each individual) can be either ``for''
or ``against''~\cite{WXL:2012}. Because of the binary (or more generally,
polarized) nature of the available data, the corresponding network
reconstruction problem is difficult. In spite of the challenge, there have
been previous efforts. For example, a compressive sensing based method
was developed to reconstruct the propagation or diffusion network of
disease spreading and to identify the source~\cite{SWFDL:2014}. In this
case, the network dynamical process is assumed to be known, e.g.,
the classical susceptible-infection-refractory (SIR) dynamics. For a variety
of binary-state dynamics, a Boltzmann machine based on the classical Ising
model can be reconstructed from the polarized data to yield the network
structure and nodal dynamics~\cite{CL:2016}, but the computational demand
is high, making the method effective but only for relatively small networks.
Quite recently, a linearization approach was proposed~\cite{LSWGL:2017} to
approximate the nodal dynamical equations that generate the binary time
series so as to convert network reconstruction into a sparse
signal optimization problem, which can then be solved by the conventional
lasso (least absolute shrinkage and selection operator) method from
statistics and machine learning. The core of this linearization approach
is to estimate the switching probabilities for a node to change from one
state to another based on properly selected and averaged strings of binary
time series, a process that requires fine adjustments of a number of
algorithmic parameters to ensure that the selected strings are neither
too special nor too similar to each other.

In this paper, we develop a statistical inference based method to reconstruct
complex network structure from binary time series. The principle of
statistical inference has recently been used in network science for tasks
such as identifying the community structures for single-layer~\cite{BKN:2011},
multilayer~\cite{BPLM:2017}, or signed~\cite{ZYLC:2017} networks, and
detecting the core-periphery structure for complex networks~\cite{ZMN:2015}.
In general, the statistical inference method has a solid mathematical
support and often can lead to robust performance. The key to structural
reconstruction is to calculate the probability for an arbitrary pair of
nodes to have a link. More importantly, it is necessary to distinguish
the probability values associated with actual links and those with
non-existent links. Accurate reconstruction demands that the two kinds
of probability values be unequivocally distinguishable. Exploiting the
expectation-maximization (EM) algorithm in statistical inference, we derive
formulas for the probabilities with the finding of a generic feature:
in all cases investigated there exists a finite gap between the
two types of probability values. Surprisingly, the appealing gap
feature is robust as it holds for a large number of combinations of the
binary state dynamics with model and real complex network structures, and
it continues to exist even when there are stochastic perturbations to the
binary time series. As a result, a threshold probability value can be
readily determined (and we provide a formula for it) to ascertain whether
there is an actual link between any pair of nodes. The final outcome is
an unprecedented high accuracy of network structural reconstruction.
Another appealing feature of our reconstruction methodology is that no
parameters are {\em a priori} assumed - all parameters can be estimated
based on the available binary data. Our reconstruction method adds a
piece into the rapidly expanding ``toolbox'' of reverse engineering of
complex dynamical networks, a problem with broad applications.

\section{Statistical inference based mathematical formulation of
reconstruction} \label{sec:math}

For a networked system hosing binary-state dynamics, at any time a node can
be in one of the two states: 0 (inactive) or 1 (active). The basic setting
under which our method is applicable is that, for the underlying
dynamical process, the probability of each node being activated at next time step is determined
\emph{only} by its active neighbors at the current step, so only transitions from inactive to active nodes are considered for the
network reconstruction. The binary dynamical process is
Markovian. Except for this condition, further details of the dynamical process
are assumed to be unknown but only the binary time series of the nodal
states are available.

\begin{figure*}
\begin{center}
\includegraphics[width=\linewidth]{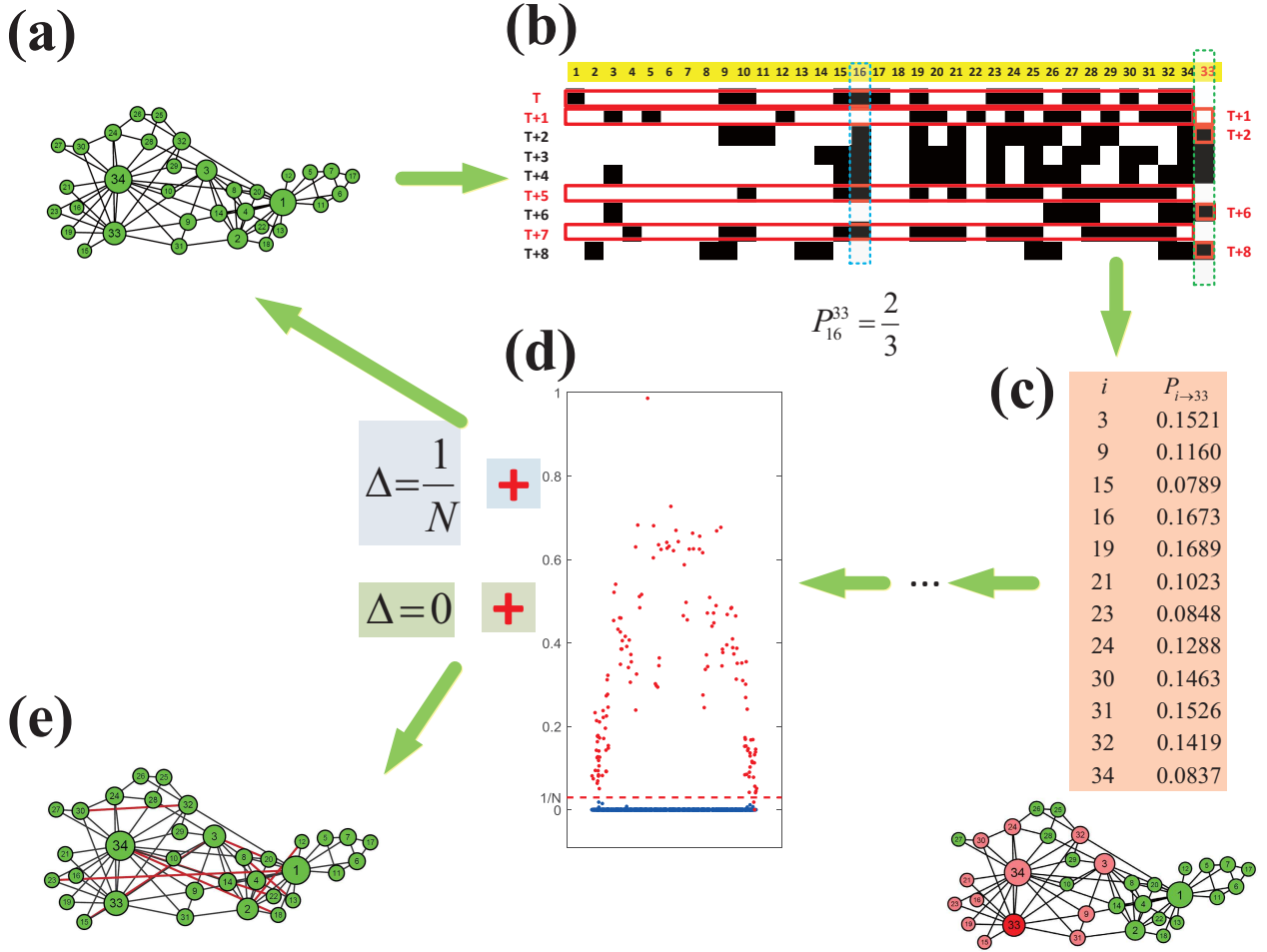}
\caption{ {\bf Schematic illustration of our binary-state network
reconstruction methodology based on statistical inference}. In this example,
the data are collected by implementing the voter model on an empirical
network - the Zachary karate-club network, where initially 30\% of the nodes
are randomly set to state $1$. (a) The actual structure of the network.
(b) The data matrix, where each row is a time string representing all
nodes' states at that time step and each column is a node's state at
different time steps. The black and blank squares denote the 1 and 0 state,
respectively. Say we wish to ascertain all neighbors of node 33 (highlighted
by the red frame), so only the strings with $\Psi^{t}_{33}=0$ and
its next string at $t+1$ are used. Each pair of useful strings are
highlighted by a frame with a different color. The quantity $P^{33}_{16}$
is the probability of the event $s_{33}(t+1)=1$ under the prior conditions
of $s_{16}(t)=1$ and $s_{33}(t)=0$. We have $P^{33}_{16}=2/3$ for this
example (highlighted by the blue frame). (c) The values of
$P_{i\rightarrow 33}$ are obtained through the EM algorithm, where only
the non-zero values of the probability are shown. The neighbors of node
$33$ in the network are shown in the lower right corner (marked by
light red color). (d) The values of $P_{i\rightarrow j}$ for each node $j$,
where the red nodes and blue points denote the actual and non-existent links,
respectively. The red dashed line represents the threshold $\Delta=1/N$
for determining whether a reconstructed value $P_{i\rightarrow j}$ can be
regarded as representing an actual link or a null link. (e) If we choose
$\Delta=0$, there are eight false links as predicted (marked by the red
links in the network). However, for $\Delta=1/N$, all actual links are
correctly inferred.}
\label{fig:method}
\end{center}
\end{figure*}

In general, if the neighbors of each node can be accurately identified,
the full topology and structure of the network can be ascertained. Consider
a network of size $N$ and $M$ time steps during the dynamical evolution.
 The available data can be represented as an $M\times N$
matrix (labeled as $S$). For example, for the illustrative network structure in
Fig.~\ref{fig:method}(a), the matrix representation of the data is the one
shown in Fig.~\ref{fig:method}(b). Let $s_i(t)$ be the state of node $i$ at
time $t$, where $s_i(t)=1$ if $i$ is active [corresponding to the black squares
in Fig.~\ref{fig:method}(b)] and $s_i(t)=0$ if $i$ is inactive [illustrated
by the blank squares in Fig.~\ref{fig:method}(b)].

Let $i \rightarrow j$ denote the event that node $i$ has a direct effect
on the state of node $j$. For example, node $i$ can spread a disease or
send a piece of information to node $j$ at time $t$. For the type of binary
dynamical processes considered, we assume that the probability for each node
to be activated is affected \emph{only} by its active neighbors. As a result,
node $i$ has a direct effect on node $j$ only when node $i$ is one of the
neighbors of node $j$. That is, the event $i \rightarrow j$ indicates whether
node $i$ is connected to node $j$, which is a property independent of time
$t$.
The conditional probability of $s_j(t+1)=1$
and $i \rightarrow j$, given $s_i(t)= 1$ and $s_j(t) = 0$, is
\begin{eqnarray} \label{eq:cond_prob}
 \nonumber P_{i\rightarrow j}^{0\rightarrow 1}&=&P[{s_j(t+1)= 1,i \rightarrow j | s_i(t)=1,s_j(t) =0}]\\
&=&P_i^{j}\cdot P_{i\rightarrow j},
\end{eqnarray}
where $P_i^{j}=P[s_j(t+1)=1|s_i(t)=1,s_j(t)=0]$ in Eq.~(\ref{eq:cond_prob})
is the probability of $s_j(t+1)=1$ under the conditions $s_i(t)=1$ and
$s_j(t)=0$, and the quantity
$P_{i\rightarrow j}=P[i \rightarrow j| s_i(t)=1, s_j(t)=0, s_j(t+1)=1]$
is the posteriori probability of $i \rightarrow j$ given $s_i(t)=1$,
$s_j(t)=0$ and $s_j(t+1)=1$.

To illustrate how the value of $P_i^{j}$ can be calculated from matrix $S$,
we consider an illustrative example. Say we know that, at time $t$=1, 5, 8
and 10, the state of node $i$ is in an active state [i.e., $s_i(t)=1$, for
$t$=1, 5, 8, 10] and state of node $j$ is in an inactive state [i.e.,
$s_j(t) = 0$, for $t$=1, 5, 8, 10]. From the matrix $S$, we have $s_j(t=2)=1$,
$s_j(t=6)=1$, $s_j(t=9)=1$, and $s_j(t=11)=0$. We get $P_i^{j}=3/4$.

Take the network in Fig.~\ref{fig:method}(a) as an example. If we wish to
infer the neighbors of node 33, we can extract some pairs of time strings, as
shown in Fig.~\ref{fig:method}(b), where each pair includes the time string
with $s_{33}(t)=0$ and its next time strings (i.e., at $t+1$). We see that
four pairs of such time strings can be extracted: $T$ and $T+1$, $T+1$ and
$T+2$, $T+5$ and $T+6$, $T+7$ and $T+8$, where each pair is highlighted by
frames with a different color. Based on these time strings, we can calculate
$P_i^{33}$ for all $i\neq j$. For example, we have $P_{16}^{33}=2/3$, as
shown in Fig.~\ref{fig:method}(b). Our goal is then to exploit statistical
inference to estimate the posteriori probability $P_{i\rightarrow j}$. Node
$i$ is a neighbor of node $j$ if $P_{i\rightarrow j}>0$, otherwise,
$P_{i\rightarrow j}=0$ if they are not connected. This analysis indicates
that the values of $P_i^{j}$ and $P_{i\rightarrow j}$ do not depend on time,
so the probability $P[{s_j(t+1)= 1,i \rightarrow j | s_i(t)=1,s_j(t) =0}]$
can simply be denoted as $P_{i\rightarrow j}^{0\rightarrow 1}$, which does
not depend on time either.

\paragraph*{Remark 1.}
For the Markovian type of dynamical processes considered, the probability
of each node's being activated is affected \emph{only} by its \emph{active}
neighbors. Other scenarios require a generalization of our method to
non-Markovian type of dynamics.

\paragraph*{Remark 2.}
To reconstruct the network structure from time series data, a necessary
condition is that the network structure should have detectable effects on
the dynamics. If the dynamical processes are independent of the network
structure, the reconstruction task is impossible. For the dynamical
processes studied, the probability of each node's being activated is
affected \emph{only} by its \emph{active} neighbors.

A non-zero value of the probability $P_i^{j}$ indicates that node $j$ is
affected by node $i$. Since the probability of each node's being activated
is determined solely by its active neighbors, a non-zero value of $P_i^{j}$
indicates an actual connection between node $i$ and $j$, which does not depend
on time. The value of $P_{i\rightarrow j}$ can be estimated once the matrix
$S$ is given, which does not depend on time either. As a result, the
probability in Eq.~(\ref{eq:cond_prob}) can be denoted as
$P_{i\rightarrow j}^{0\rightarrow 1}$. From Eq.~(\ref{eq:cond_prob}), we
see that, if node $j$ is not activated at time $t_m$, the expected number
of node $j$ being activated by its neighbors at $t_m+1$ is given by

\begin{eqnarray} \label{eq:ENI}
\nonumber  E_j^{t_m + 1}& = & \sum_{i\neq j}
P_{i\rightarrow j}^{0\rightarrow 1}\Psi^{t_m}_{i}
+\varepsilon_j\\&=&\sum_{i\neq j} P_{i\rightarrow j}\cdot
P_i^{j}\Psi^{t_m}_{i}+\varepsilon_j,
\end{eqnarray}
 where $\Psi^{t_m}_{i}=1$ when node $i$ was activated at
time $t_m$, otherwise, $\Psi^{t_m}_{i}=0$. $\varepsilon_j$ characterizes the stochastic influence (noise)
on node $j$.

Note that, due to the errors from the collected data and the assumptions
used in the development of the method (e.g., the assumption of the Poisson
distribution), it is necessary to consider the presence of noise perturbation
in Eq.~(\ref{eq:ENI}). While different types of noise can be considered,
additive noise facilitates both computation and analysis, as done in previous
works (e.g., Ref.~\cite{LSWGL:2017}).

To simplify notation, we let $\Theta$ denote the quantities
$P_{i\rightarrow j}$ and $\varepsilon_j$. To derive analytically an
EM estimation, we assume that the relevant probability distributions are
Poisson~\cite{BKN:2011,BPLM:2017}.
The reason is that Poisson distribution can be generally used to
characterize the probability of a given number of events occurring in
a fixed interval of time. It is thus natural to use Poisson distribution
to describe the \emph{times} that node $i$ being activated. As in
Refs.~\cite{Newman:2016,NR:2016,KN:2011,BPLM:2017}, using the Poisson
distribution can make feasible mathematical analysis and computations with
the EM algorithm (described below). We note that, with any assumption of
the probability distribution, errors are inevitable. For example, the value
of $P_{i\rightarrow j}$ may be slightly larger than zero even though node
$i$ is not a neighbor of node $j$. To reduce such errors, a remedy is to
set a cutoff threshold to determine if $P_{i\rightarrow j}>0$ indicates
an actual link or it is simply an error.

The probability $\Psi_j$ can then be expressed as
\begin{eqnarray}\label{eq:ML}
\nonumber &&P\left( {{{\left\{ {\Psi _j^{{t_m} + 1}} \right\}}_{m = 1, \cdots ,M}}\left| {\Theta ,{{\left\{ {\Psi _i^{{t_m}}} \right\}}_{m = 1, \cdots ,M;i = 1, \cdots ,N}}} \right.} \right) = \\&&
\prod\limits_{m,\Psi _j^{{t_m}} = 0} {\frac{{{e^{ - E_j^{{t_{m}+1}}}}{{\left( {E_j^{{t_{m}+1}}} \right)}^{\Psi _j^{{t_{m}+1}}}}}}{{\Psi _j^{{t_{m}+1}}!}}}
\end{eqnarray}

Next, we exploit the EM method to maximize the likelihood
Eq.~(\ref{eq:ML}) so that the model parameters $\Theta$ can be estimated
from the binary data.
The EM algorithm is general for finding the maximum likelihood estimate in
latent variable models, which contains two steps. For the E-Step, one ``fills
in'' the latent variables using the posterior probability and, for the
M-Step, one maximizes the expected complete logarithmic likelihood with respect
to the complete posterior distribution. Jensen's inequality is a key tool in
the M-step for generating the EM objective function. A comprehensive
explanation of the principle of EM algorithm can be found in
Ref.~\cite{DLR:1977}. The algorithm has also been widely used in network
structure reconstruction, e.g., in
Refs.~\cite{Newman:2016,NR:2016,KN:2011,BPLM:2017}. It is convenient to
maximize the logarithm of the likelihood:
\begin{eqnarray} \label{eq:ML_log}
\nonumber L\left( \Theta  \right)&&= \sum\limits_{m,~\Psi _j^{t_m} = 0} {\left[ {\Psi _j^{t_m + 1}
\log\left( {\sum\limits_{i \ne j} {{P_{i \to j}}P_i^j\Psi _i^{t_m}}
+ {\varepsilon_j}} \right)} \right.} \\
&& - \left. {\left( {\sum\limits_{i \ne j} {{P_{i \to j}}P_i^j\Psi _i^{t_m}}
+ {\varepsilon _j}} \right)} \right],
 \end{eqnarray}


$\Psi_j^{{t_m+1}}!\equiv1$ since $\Psi_j^{{t_m+1}}$ equals 0 or 1 in this work, therefore we have omitted the terms $\Psi_j^{{t_m+1}}!$. Using Jensen's inequality~\cite{Needham:1993},
we obtain
\begin{eqnarray} \label{eq:Jensen}
\nonumber&&\log \left( {\sum\limits_{i\ne j} {{P_{i \to j}}P_i^j\Psi_i^{t_m}}
+ {\varepsilon _j}} \right) \\&&
\nonumber= \log \left( {\sum\limits_{i \ne j}
{\rho_i^{t_m}\frac{{{P_{i \to j}}P_i^j\Psi_i^{t_m}}}{{\rho_i^{ {t_m} }}}}
+ \rho_{\varepsilon}^{t_m}\frac{{\varepsilon_j}}{\rho_{\varepsilon}^{t_m}}} \right)\\
&& \nonumber\ge \sum\limits_{i \ne j} {\rho_i^{t_m}\log \frac{{{P_{i \to j}}
P_i^j\Psi_i^{t_m}}}{{\rho_i^{t_m}}}} + \rho_{\varepsilon}^{t_m}
\log\frac{{{\varepsilon _j}}}{{\rho _\varepsilon ^{t_m}}}\\
&& \nonumber= \sum\limits_{i \ne j} {\rho_i^{t_m}\log {P_{i \to j}}
P_i^j\Psi _i^{t_m} + } \rho_{\varepsilon}^{t_m}\log {\varepsilon_j}\\ &&
- \sum\limits_{i \ne j} {\rho_i^{t_m}\log\rho_i^{t_m} - } \rho_{\varepsilon}^{t_m}
\log\rho_{\varepsilon}^{t_m},
\end{eqnarray}
where
\begin{eqnarray} \label{eq:rho_T}
\rho_i^{t_m} = \frac{{{P_{i \to j}}P_i^j\Psi_i^{t_m}}}{{\sum\limits_{i' \ne j}
{{P_{i' \to j}}P_{i'}^j\Psi_{i'}^{t_m}} + {\varepsilon_j}}}
\end{eqnarray}
and
\begin{eqnarray} \label{eq:rho_eps}
\rho_{\varepsilon}^{t_m} = \frac{\varepsilon_j}{{\sum\limits_{i'\ne j}
{{P_{i'\to j}}P_{i'}^j\Psi_{i'}^{t_m}} + {\varepsilon _j}}}.
 \end{eqnarray}
To find a maximum likelihood solution of Eq.~(\ref{eq:ML_log}), we seek
to maximize the following quantity:
\begin{eqnarray} \label{eq:L}
\nonumber &&L\left({\Theta,\rho}\right) =
\sum\limits_{m,~\Psi_j^{t_m} = 0} {\sum\limits_{i \ne j}
{\left({\Psi_j^{{t_m}+1}\rho_i^{t_m}\log {P_{i \to j}}P_i^j\Psi_i^{t_m}}\right.} } \\
&&\left. {-\Psi_j^{t_m+1}\rho_i^{t_m}\log \rho_i^{t_m} - {P_{i \to j}}P_i^j\Psi_i^{t_m}}
\right)+\\&& \nonumber
 \sum\limits_{m,~\Psi_j^{t_m} = 0} {\left[{\Psi_j^{{t_m}+1}\rho_{\varepsilon}^{t_m}
\log {\varepsilon_j} - \Psi_j^{{t_m}+1}\rho_{\varepsilon}^{t_m}
\log\rho_{\varepsilon}^{t_m} - {\varepsilon_j}} \right]}
\end{eqnarray}
with respect to $\Theta$ and $\rho$. Calculating the partial derivative of
$L(\Theta,\rho)$ with respect to $P_{i\rightarrow j}$ and $\varepsilon_j$
and setting them to be zero, we have
\begin{eqnarray} \label{eq:LP}
\frac{{\partial L\left({\Theta,\rho}\right)}}{{\partial {P_{i\to j}}}}
= \sum\limits_{m,~\Psi_j^{t_m} = 0} {\left( {\frac{{\Psi_j^{{t_m} + 1}\rho_i^{t_m}}}
{{{P_{i\to j}}}} - P_i^j\Psi _i^{t_m}}\right)} =0
\end{eqnarray}
and
\begin{eqnarray} \label{eq:Leps}
\frac{{\partial L\left({\Theta,\rho} \right)}}{{\partial {\varepsilon _j}}}
= \sum\limits_{m,~\Psi_j^{t_m} = 0}{\left({\frac{{\Psi_j^{{t_m}+1 }
\rho_{\varepsilon}^{t_m}}}{{{\varepsilon_j}}}-1} \right)} = 0,
\end{eqnarray}
which give
\begin{eqnarray}\label{eq:Pij}
P_{i\to j} = \frac{{\sum\limits_{m,~\Psi_j^{t_m}=0}
{\left( {\Psi_j^{{t_m}+1}\rho_i^{t_m}} \right)}}}
{{\sum\limits_{m,~\Psi_j^{t_m}=0}{\left({P_i^j\Psi _i^{t_m}}\right)}}}
\end{eqnarray}
and
\begin{eqnarray} \label{eq:eps_j}
\varepsilon_j = \frac{{\sum\limits_{m,~\Psi_j^{t_m} = 0}
{\left({\Psi_j^{{t_m}+1}\rho_i^{t_m}} \right)} }}
{{\sum\limits_{m,~\Psi_j^{t_m} = 0}{\left(1 \right)} }},
\end{eqnarray}
respectively.

Equations~(\ref{eq:rho_T}), (\ref{eq:rho_eps}), (\ref{eq:Pij}) and
(\ref{eq:eps_j}) constitute our method. From the initial conditions of
$P_{i \to j}$ and $\varepsilon_j$, we can iterate these equations until
convergence is achieved. Since a single iterative process does not ensure
global optimization, we carry out the above iteration process several
times and choose the relevant values that give the maximum of the
quantity in Eq.~(\ref{eq:ML_log}). As an example, Fig.~\ref{fig:method}(c)
shows the value of $P_{i\rightarrow 33}$ (only $P_{i\rightarrow 33}>0$ is
shown) calculated from the iterative process. Similarly, the values of
$P_{i\rightarrow j}$ for all the nodal pairs can be calculated, as
shown in Fig.~\ref{fig:method}(d), where the red and blue dots denote the
actual and non-existent links, respectively. Theoretically, node $i$ is a
neighbor of node $j$ if $P_{i\rightarrow j}>0$ with the threshold
value $\Delta=0$. However, the simple choice of $\Delta=0$ will lead to
error due to the uncertain factors. For example, as shown in
Fig.~\ref{fig:method}(e), there are eight false links (represented by the
red lines). In this case, it is necessary to choose a non-zero threshold
for each node to eliminate reconstruction error. For instance, by
setting $\Delta=1/N$ for all nodes in Fig.~\ref{fig:method}, we can
reconstruct the original network with zero error.

An explanation is in order. It is often difficult to directly maximize the
formula in Eq.~(\ref{eq:ML_log}). We can first use Jensen's inequality
to get the lower bound of the formula at $\Theta$, which is denoted by
$L^{-}(\Theta)=\sum\limits_{i \ne j} {\rho_i^{t_m}\log {P_{i \to j}}
P_i^j\Psi _i^{t_m} + } \rho_{\varepsilon}^{t_m}\log {\varepsilon_j}
- \sum\limits_{i \ne j} {\rho_i^{t_m}\log\rho_i^{t_m} - }
\rho_{\varepsilon}^{t_m} \log\rho_{\varepsilon}^{t_m}$
- the last term of Eq.~(\ref{eq:Jensen}). We initialize the parameter
$\Theta_1$, e.g., by setting
$P_{1\rightarrow j}=P_{j-1\rightarrow j}=P_{j+1\rightarrow j}
= \ldots,P_{N\rightarrow j}=\varepsilon_j=1/N$. We can show that the
equality conditions in Eq.~(\ref{eq:L}) are satisfied when the conditions
in Eqs.~(\ref{eq:rho_T}) and (\ref{eq:rho_eps}) are met. We thus have
$L^{-}_{\Theta_1}(\Theta_1)=L(\Theta_1)$, where $L^{-}_{\Theta_1}(\Theta)$
denotes the lower bound function of $L(\Theta)$ at $\Theta_1$, so
$L^{-}_{\Theta_1}(\Theta_1)$ indicates the value of $L^{-}_{\Theta_1}(\Theta)$
at $\Theta=\Theta_1$. Further, by maximizing $L^{-}_{\Theta_1}(\Theta)$, we
obtain a new maximum point $\Theta_2$:
\begin{displaymath}
L(\Theta_2)\geq L^{-}_{\Theta_1}(\Theta_2)\geq L^{-}_{\Theta_1}(\Theta_1)
=L(\Theta_1),
\end{displaymath}
meaning that $\Theta_2$ is a better solution than $\Theta_1$.

We also note that the initial conditions of $P_{i\rightarrow j}$ and
$\varepsilon_j$ can be chosen in different ways. For example, we can set
$P_{1\rightarrow j}=P_{j-1\rightarrow j}=P_{j+1\rightarrow j}
=\ldots,P_{N\rightarrow j}=\varepsilon_j=1/N$. The quantities $\rho_i^{t_m}$
and $\rho_{\varepsilon}^{t_m}$ in Eqs.~(\ref{eq:rho_T}) and
(\ref{eq:rho_eps}) can be calculated, guaranteeing the equality
condition in Eq.~(\ref{eq:L}). Then, Eqs.~(\ref{eq:Pij}) and
(\ref{eq:eps_j}) can be calculated. Iterating the above process leads to
a local optimal solution. The value of the
likelihood function at the next time step is better than that at the last
step. Since the convergence of the EM algorithm has been confirmed in
many previous works, we can stop the iteration process when the value of
the likelihood function is stable or the fluctuations are smaller than a
given threshold value. While one round of the iteration may yield a local
rather than a global optimal solution, we can choose different sets of
initial values to carry out different rounds of iteration and choose the best
solution.

We remark that errors in the collected
data and uncertainties in the assumption of the Poisson distribution can be
modeled by noise perturbation. The simple choice of $\Delta$ will lead to
small errors. While errors cannot be completely eliminated by increasing the
value of $M$, the gap that is key to distinguishing actual from non-existent
links will be enlarged. Figure~\ref{fig:method} shows that the accuracy of
reconstruction can be improved if we set $\Delta=1/N$ for all nodes.

\section{Performance characterization and demonstration}
\label{sec:performance}

\subsection{Local and global performance indicators} \label{subsec:indicators}

We use a number of indicators to characterize the local and global
performance of our reconstruction methodology.

\paragraph*{AUROC and AUPR - local performance indicators.}
The AUROC (area under receiver operating characteristic,$\delta_{AUROC}$) and AUPR (area under
precision-recall,$\delta_{AUPR}$) curves are standard local (node-wise) performance indicators
used widely in signal processing and computer science~\cite{DG:2006},
which can be calculated for each node in the network. The average values
over all the nodes can then be used to characterize the reconstruction
performance for the whole network. To define AUROC and AUPR, it is
necessary to calculate three basic quantities: TPR (true
positive rate,$R_{TP}$), FPR (false positive rate,$R_{FP}$), and Recall. In particular,
TPR is defined as
\begin{equation} \label{eq:TPR}
\mbox{$R_{TP}$}(l)=\frac{\mbox{$P_T$}(l)}{P},
\end{equation}
where $l$ is the cutoff in the list of reconstructed links, ${P_T}(l)$
is the number of true positives in the top $l$ predictions in the link
list, and $P$ is the number of positives. FPR is given by
\begin{equation} \label{eq:FPR}
\mbox{$R_{FP}$}(l)=\frac{\mbox{$P_F$}(l)}{Q},
\end{equation}
where ${P_F}(l)$ is the number of false positive in the top $l$
predictions in the link list, and $Q$ is the number of negatives in the
gold standard. The reconstruction precision can be defined as
\begin{equation} \label{eq:precision}
\mbox{$\delta_{Precision}$}(l)=\frac{\mbox{$P_T$}(l)}{\mbox{$P_T$}(l)+\mbox{$P_F$}(l)}
=\frac{\mbox{$P_T$}(l)}{l}.
\end{equation}
The measure Recall is defined as
\begin{equation} \label{eq:recall}
\mbox{$\delta_{Recall}$}(l)=\mbox{$R_{TP}$}(l)=\frac{\mbox{$P_T$}(l)}{P}.
\end{equation}
Varying the value of $l$ from $0$ to $N$, we plot two sequences of points:
[$R_{FP}(l),R_{TP}(l)$] and [$\delta_{Recall}(l),\delta_{Precision}(l)$].
The areas under the two curves give the values of AUROC and AUPR, respectively.
For the case of zero error in reconstruction where all the actual links
have been predicted, we have $\delta_{AUROC}=1$ and $\delta_{AUPR}=1$. In the worse case scenario
where the predicted links are completely random (so that the reconstruction
tasks fails entirely), we have $\delta_{AUROC}=0.5$ and $\delta_{AUPR}=P/2N$.

\begin{figure}
\centering
\includegraphics[width=\linewidth]{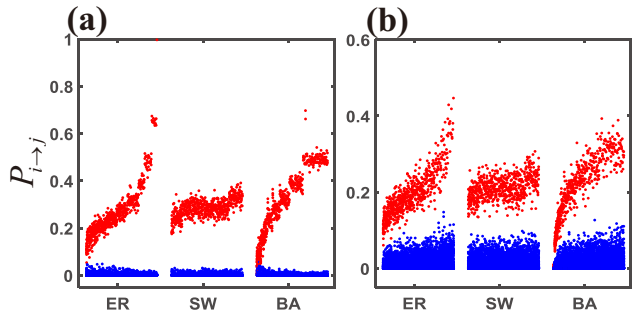}
\caption{ {\bf Demonstration of placement of threshold probability value
for calculating the global performance indicator F1}. For combinations of
two types of binary-state dynamics [voter dynamics in (a) and Kirman dynamics
in (b)] and three complex network topologies, the values of
$P_{i\rightarrow j}$ for $i\neq j$ ($i=1,2,\cdots,N$) for each node in the
network are shown. The result for a node corresponds to a column above
the x-axis consisting of $N-1$ number of points. The red nodes and blue
points denote the actual and non-existent links, respectively.
Three model networks (ER, SW and SF networks) are used. All networks have
$N=100$ nodes and average degree $\langle k\rangle=6$. The length of the
binary time series is $M=15000$.}
\label{fig:PFM}
\end{figure}

\paragraph*{F1 score - a global performance indicator.}
Higher values of AUROC and AUPR \emph{only} demonstrate that the prediction
of the actual links are better than that for the non-existent links, but
do not give the number of actual links in the network. These local
measures do not indicate whether a specific link has been correctly inferred.
To determine whether a reconstructed probability value (i.e.,
$P_{i\rightarrow j}$) corresponds to an actual or a null link, it is
necessary to set a threshold $\Delta$ for each node.
Figure~\ref{fig:PFM} shows, for each node, the value of $P_{i\rightarrow j}$
for $i\neq j$ ($i=1,2,\cdots, N$) in three model networks:
random network (ER)~\cite{ER:1960},
scale-free network (BA or SF)~\cite{BA:1999}, and small-world (SW)
network~\cite{WS:1998}. The dynamical processes are Voter dynamics in
Fig.~\ref{fig:PFM}(a) and Kirman dynamics in Fig.~\ref{fig:PFM}(b)),
respectively. (The details of these two processes, together with other
six other types, are given in Appendix.)
We see that, for node $j$, there exists a gap dividing
the values of $P_{i\rightarrow j}$ for $i\neq j$ ($i=1,2,\cdots,N$).
It is thus reasonable to place a threshold $\Delta_j$ in the gap for
node $j$ to determine whether a value of $P_{i\rightarrow j}$ can be regarded
as representing an actual links (red points, $P_{i\rightarrow j}>\Delta_j$)
or a non-existent link (blue points, $P_{i\rightarrow j}<\Delta_j$). In so
doing, we obtain the nonzero values $P_{i\rightarrow j}>0$ for $i\neq j$
and re-rank them in a descending order, denoted
as $P'_l$ ($l=1, 2,\cdots$).

It is important to choose a proper threshold $\Delta_j$ for each node $j$. From
Fig.~\ref{fig:PFM}, we see that there is a gap, which can be used to separate
the actual from the non-existent links. Computationally, it is necessary to set
a threshold for the task. We consider two different scenarios. First, suppose
that a sequence of the values of $P_{i\rightarrow j}$ is 0.8, 0.7, 0.6, 0.01,
and 0.0001. In this case, the threshold can be set between the values of
0.6 and 0.01 through the maximum value of $P'_{l}-P'_{l+1}$. However, the threshold value is between 0.01
and 0.0001 when using $P'_{l}/P'_{l+1}$. For this scenario, the former choice of the threshold value is
more reasonable than the latter. Second, for a different sequence, such
as 0.2, 0.1, 0.09 and 0.0001, through $P'_{l}-P'_{l+1}$ we find a threshold
value between 0.2 and 0.1. However, through $P'_{l}/P'_{l+1}$, we get
a threshold value between 0.09 and 0.0001. For this scenario, the latter
case is more reasonable. Combining the two cases, we define the threshold
$\Delta_j$ for node $j$ as
\begin{equation} \label{eq:threshold}
\Delta_j=\arg \max_{l}\large[\frac{P_l'}{P'_{l+1}}(P'_l-P'_{l+1})\large].
\end{equation}
With the threshold value so determined, we can ascertain, for any pair of
nodes in the network, whether there is an actual link. The $\mbox{F1}$ score is given
by~\cite{Power:2011};
\begin{equation} \label{eq:F1}
\mbox{F1}=\frac{2\delta_{Precision}\delta_{Recall}}
{\delta_{Precision} + \delta_{Recall}},
\end{equation}
where $\delta_{Precision}=P_T/(P_T+P_F)$ and
$\delta_{Recall}=P_T/(P_T+N_F)$ respectively. The quantities
$P_T$, $N_F$, $P_F$ and $N_T$ denote the true
positive, false negative, false positive and true negative. The condition
$\mbox{F1}=1$ indicates that the reconstructed links perfectly match with
those in the original network.

Another global indicator, denoted by $\mbox{ERR}$($R_{ER}$), is defined as the ratio
of the number of erroneous links (false positive and false negative) to
the number of links of the true network. Namely,
\begin{equation} \label{eq:ERR}
R_{ER}=\frac{N_F+P_F}
{P_T+N_F}.
\end{equation}

\subsection{Reconstruction performance with model and real networks}
\label{subsec:RealN}

\begin{table*}
\centering
\footnotesize
\caption{ {\bf Local reconstruction performance with model and real networks}.
Values of AUROC and AUPR for various dynamics on a variety of model and
empirical (real) networks. The parameters in the dynamical models are
described in Appendix. The size and average degree of the
three types of model complex networks (ER, BA, and SW) are $N = 500$ and
$\langle k\rangle=6$. The length of the binary data string is $M=50000$
for $N=500$, $M=15000$ for $N<500$, and $M=100000$ for $N>1000$. The
largest values AUROC and AUPR for each case is highlighted in bold. For
comparison, the corresponding AUROC and AUPR values from the recent lasso
method~\cite{LSWGL:2017} are also shown.}
\begin{tabular}{c|ccccccccc}
\hline
\multicolumn{2}{c|}{AUROC/AUPR}  &Voter  &Kirman &Ising  &SIS &Game &Language &Threshold &Majority\\ \hline
\multirow{2}{*}{Karate}
                   &lasso &0.980/0.971 &0.990/0.959 &0.997/0.997 &0.954/0.946 &0.993/0.992 &0.961/0.926 &0.995/0.996 &0.997/0.996 \\
                   &EM &\textbf{0.999/0.999} &\textbf{1.000/1.000} &\textbf{1.000/1.000} &\textbf{0.983/0.982}
                    &\textbf{1.000/1.000} &\textbf{0.998/0.998} &\textbf{1.000/1.000} &\textbf{1.000/1.000} \\ \hline
 \multirow{2}{*}{Dolphins}
                   &lasso &0.974/0.917 &0.996/0.984 &0.999/0.997 &0.981/0.941 &0.996/0.988 &0.987/0.945 &\textbf{1.000/1.000} &0.998/0.992 \\
                   &EM &\textbf{1.000/1.000} &\textbf{1.000/1.000} &\textbf{1.000/1.000} &\textbf{0.998/0.993}
                    &\textbf{1.000/1.000} &\textbf{1.000/0.999} &{1.000/0.999} &\textbf{1.000/1.000} \\ \hline
\multirow{2}{*}{Polbooks}
                   &lasso &0.967/0.865 &0.984/0.912 &0.989/0.968 &0.896/0.801 &0.974/0.926 &0.951/0.851 &\textbf{1.000/0.999} &0.983/0.943 \\
                   &EM &\textbf{1.000/0.999} &\textbf{1.000/1.000} &\textbf{1.000/0.999} &\textbf{0.940/0.864}
                    &\textbf{0.994/0.991} &\textbf{0.991/0.975} &{0.998/0.998} &\textbf{0.999/0.997} \\ \hline
\multirow{2}{*}{Football}
                   &lasso &0.959/0.812 &0.991/0.949 &0.991/0.950 &0.928/0.711 &0.986/0.920 &0.927/0.703 &\textbf{1.000/1.000} &0.987/0.927 \\
                   &EM &\textbf{1.000/1.000} &\textbf{1.000/1.000} &\textbf{1.000/1.000} &\textbf{0.996/0.973}
                    &\textbf{0.999/0.998} &\textbf{0.999/0.994} &\textbf{1.000/1.000} &\textbf{1.000/0.999} \\ \hline
\multirow{2}{*}{Email}
                   &lasso &0.943/0.781 &0.655/0.331 &0.971/0.808 &0.789/0.607 &0.968/0.860 &0.923/0.622 &1.000/0.998 &0.965/0.723 \\
                   &EM &\textbf{1.000/1.000} &\textbf{0.955/0.799} &\textbf{1.000/1.000} &\textbf{0.977/0.893}
                    &\textbf{0.999/0.997} &\textbf{0.999/0.990} &\textbf{1.000/1.000} &\textbf{1.000/1.000} \\ \hline
 \multirow{2}{*}{ER(500)}
                   &lasso &0.999/0.975 &0.988/0.784 &0.998/0.974 &0.994/0.972 &0.999/0.979 &0.977/0.751 &\textbf{1.000/1.000} &0.996/0.929 \\
                   &EM &\textbf{1.000/1.000} &\textbf{1.000/1.000} &\textbf{1.000/1.000} &\textbf{1.000/0.997}
                    &\textbf{1.000/1.000} &\textbf{1.000/1.000} &\textbf{1.000/1.000} &\textbf{1.000/1.000} \\ \hline
\multirow{2}{*}{SW(500)}
                   &lasso &\textbf{1.000/1.000} &0.992/0.838 &1.000/0.998 &\textbf{1.000/1.000} &1.000/0.998 &0.997/0.930 &\textbf{1.000/1.000} &0.998/0.937 \\
                   &EM &\textbf{1.000/1.000} &\textbf{1.000/1.000} &\textbf{1.000/1.000} &\textbf{1.000/1.000}
                    &\textbf{1.000/1.000} &\textbf{1.000/1.000} &\textbf{1.000/1.000} &\textbf{1.000/1.000} \\ \hline
\multirow{2}{*}{BA(500)}
                   &lasso &0.996/0.953 &0.940/0.697 &0.994/0.963 &0.968/0.926 &0.989/0.946 &0.978/0.861 &1.000/0.998 &0.994/0.944 \\
                   &EM &\textbf{1.000/0.999} &\textbf{0.992/0.971} &\textbf{1.000/1.000} &\textbf{0.983/0.949}
                    &\textbf{0.998/0.997} &\textbf{0.998/0.992} &\textbf{1.000/1.000} &\textbf{1.000/1.000} \\ \hline
\end{tabular}\label{table1}
\end{table*}

We consider eight types of binary-state dynamical processes as studied recently
in Ref.~\cite{LSWGL:2017} with the lasso method. For the
network structures, we use three types of model complex networks (ER, SF,
and WS) and a number of empirical networks as described in Appendix. In
Tab.~\ref{table1}, we compare the performance of our EM algorithm with that of
the lasso method under the same setting. We see that the performances of
the two methods for the threshold dynamics are almost identical as both
exhibit nearly perfect values of AUROC and AUPR (almost $100\%$). However,
for the other seven types of binary-state dynamics in combination with
different network structures (model or empirical networks), our EM based
reconstruction method yields results that are more accurate than those
with the lasso method. The value of F1 scores from our method for various
combinations of network structures and binary-state dynamics are summarized
in Tab.~\ref{table2}, where we see that the values of F1 score in most cases
are close to unity, indicating accurate reconstruction performance.
Since the lasso method does not rely on any threshold value for each
node~\cite{LSWGL:2017}, it is not feasible to compare performance in
terms of the F1 score.

\begin{table*}
\centering
\small
\caption{
{\bf Characterization of global performance of proposed statistical
inference based reconstruction method}. Listed are the values of F1 score
and ERR for various combinations of binary dynamics and networks (model and
empirical), where the threshold $\Delta_j$ for each node is determined
according to Eq.~(\ref{eq:threshold}). Other parameters are the same as in
Tab.~\ref{table1}.}
%
%
%
%
%
%
%

\begin{tabular}{ccccccccc} \hline
 F1/ERR &Voter  &Kirman &Ising  &SIS &Game &Language &Threshold &Majority\\ \hline
         Karate &0.994/0.013 &1.000/0.000 &1.000/0.000 &0.981/0.039 &0.994/0.013 &1.000/0.000 &0.947/0.103 &1.000/0.000 \\

         Dolphins &1.000/0.000 &1.000/0.000 &0.997/0.006 &0.984/0.031 &0.994/0.013 &1.000/0.000 &1.000/0.000 &0.994/0.013 \\

         polbooks &0.986/0.027 &1.000/0.000 &0.990/0.020 &0.867/0.254 &0.972/0.054 &0.960/0.077 &0.986/0.027 &0.972/0.057 \\

         Football &1.000/0.000 &0.999/0.002 &0.999/0.002 &0.844/0.277 &0.992/0.015 &0.941/0.116 &0.963/0.072 &0.992/0.016 \\

         Email &0.998/0.004 &0.712/0.531 &0.998/0.004 &0.853/0.265 &0.984/0.031 &0.943/0.108 &0.995/0.010 &1.000/0.001 \\

         ER(500) &1.000/0.000 &1.000/0.000 &1.000/0.000 &0.998/0.005 &1.000/0.000 &1.000/0.000 &1.000/0.000 &1.000/0.000 \\

         SW(500) &1.000/0.000 &1.000/0.000 &1.000/0.000 &1.000/0.000 &1.000/0.000 &1.000/0.000 &1.000/0.000 &1.000/0.000 \\

         BA(500) &0.998/0.006 &0.930/0.142 &0.997/0.008 &0.968/0.064 &0.993 /0.016 &0.984/0.033 &0.995/0.011 &0.999/0.003 \\ \hline
\end{tabular}\label{table2}
\end{table*}

\begin{figure}
\centering
\includegraphics[width=\linewidth]{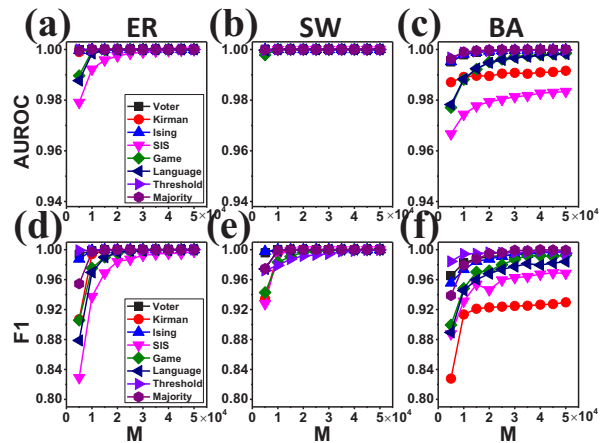}
\caption{ {\bf Dependence of reconstruction performance on data length}.
For the eight types of binary-state dynamics implemented on ER, SW, and SF
networks, AUROC (a-c) and F1 score (d-f) versus $M$, the length of the
binary time series, for ER (left panel), SW (central panel), and SF (right
panel) networks. All networks have $N = 500$ nodes with the average degree
$\langle k \rangle = 6$. The parameters for each type of dynamics
are described in Appendix. For the large number of combinations of binary
dynamical processes and network topologies, both the local (AUROC) and
global (F1 score) performance measures approach almost the highest possible
values when $M$ becomes sufficiently large.}
\label{fig:DoM}
\end{figure}

Figure~\ref{fig:DoM} shows, for the model networks, the dependence of the
values of AUROC and F1 score on $M$, the length of the binary time series,
where we see that, in all cases, AUROC approaches a stable and
large (e.g., $>0.97$) value for $M\approx 25000$. The values of F1 score are
also large (e.g., $> 0.92$). In terms of the network topology, the highest
performance is achieved for the SW, followed by ER and then SF networks.
The intuitive reason for the relatively inferior performance with SF networks
lies in the difficulty to infer the neighbors of hub or high degree nodes.

\begin{figure}
\centering
\includegraphics[width=\linewidth]{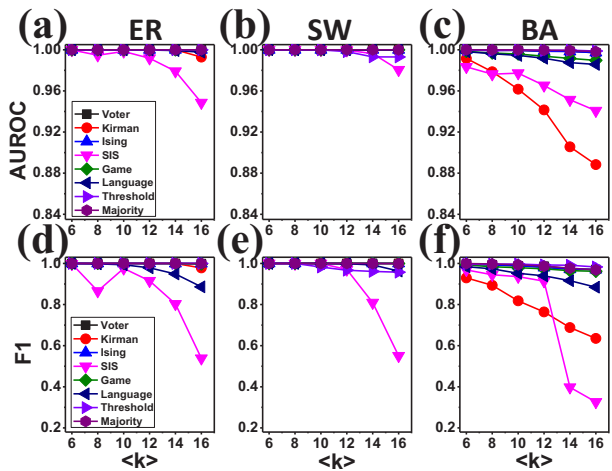}
\caption{ {\bf Effect of increasing the network average degree on
reconstruction performance}. For the eight types of binary-state dynamics
implemented on ER, SW, and SF networks, AUROC (a-c) and F1 score (d-f)
versus the average degree $\langle k\rangle$, for ER (left panel), SW
(central panel), and SF (right panel) networks. All networks have $N = 500$
nodes and the length of the binary time series is fixed at $M=50000$.
In general, the reconstruction accuracy tends to decrease as the average
degree becomes larger.}
\label{fig:DoK}
\end{figure}

Figure~\ref{fig:DoK} shows results, for the combinations of eight binary
dynamical processes and the three distinct complex network topologies,
the dependence of the reconstruction performance on the network average
degree $\langle k\rangle$. We note a decreasing trend in the reconstruction
accuracy as the average degree becomes larger. The reason is that an increase
in $\langle k\rangle$ demands more links to be predicted, leading to a decrease
in the reconstruction accuracy if the data length is not increased accordingly.
Another phenomenon is that, except for the SIS and Kirman dynamics, the
average degree does not have an appreciable effect on the reconstruction
accuracy. The heuristic reason of the relatively stronger dependence
of the reconstruction performance on the average degree for the SIS and
Kirman dynamics is that, for these two types of dynamics, the probability
of being activated is proportional to the number of active neighbors $m$
rather than the density of the active neighbors, $m/k$. As a result,
increasing the average degree will expedite the dynamical
propagation of the ``virus'' or information, leading to most nodes being
activated in relatively short time. From the standpoint of reconstruction,
this is damaging due to lack of sufficient information about the time
evolution of the underlying dynamics. To improve the reconstruction
performance, one can reduce the transmission rate $\lambda$ in the SIS
process and the transmission rate $c_1+md$ in the Kirman dynamics. On the
contrary, for other six types of binary-state dynamics, for a large
average degree value, the probability of being activated is not
significantly increased due to its dependence on the density $m/k$ (not on
$m$ itself), so the slow pace of the dynamical evolution on the networks
persists and, consequently, there is still sufficient amount of information
required for the reconstruction task.

\begin{figure}
\centering
\includegraphics[width=\linewidth]{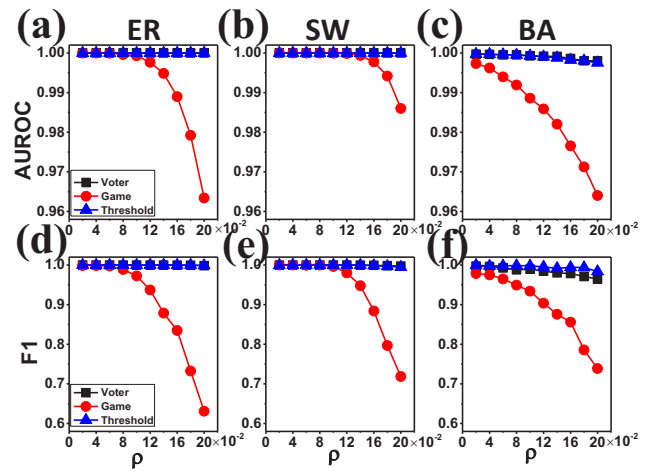}
\caption{ {\bf Effect of stochastic disturbance on reconstruction}.
The local (AUROC, a-c) and global (F1 score, d-f) performance indicators
versus $\rho$, the fraction of randomly flipped binary states in the data,
for the voter (blue squares), game (red circles), and threshold (blue
up triangles) models, for ER (left panel), SW (central panel), and SF
(right panel) networks. All networks have the size $N = 500$ with the
average degree $\langle k\rangle=6$. The length of the binary time series
is $M=50000$.}
\label{fig:Noise}
\end{figure}

Finally, we demonstrate the robustness of our EM algorithm against stochastic
disturbance. Specifically, we randomly flip a fraction $\rho$ of the binary
states among the total number $MN$ of states and calculate the values of
AUROC and F1 score versus $\rho$ for various combinations of the dynamics
and network topology. The results are shown in Fig.~\ref{fig:Noise}. From
the top panel, we see that the values of AUROC are larger than $0.96$ even
when 20\% of the states are flipped, which are more robust than those with
the lasso method (e.g., Tab.~$3$ in Ref.~\cite{LSWGL:2017}). We also see
that the reconstruction performances with the voter and threshold dynamics
are relatively more robust to stochastic perturbations than those with
the other six types of dynamical processes. A possible reason is that, in
the game dynamics, each node's payoff depends sensitively on the neighbors'
states. If one neighbor's state is flipped, there can be a dramatic change
in the payoff, affecting directly its strategy (cooperation or defection)
and consequently the reconstruction accuracy.

\section{Discussion} \label{sec:discussion}

In physics and mathematics, the various inverse problems to infer the
internal structure or ``gears'' of the underlying system based on
observations are always challenging. For complex networked systems,
recent years have witnessed the development of various frameworks and
methodologies to address the inverse or reverse-engineering
problem~\cite{WLG:2016}, leading to the gradual establishment of a ``toolbox''
of network and dynamics reconstruction algorithms to deal with a variety of
specific tasks. This work adds another piece into this toolbox: a
statistical inference based method specifically designed to address the
network reconstruction problem for binary dynamical processes without
requiring any {\em a priori} knowledge about the switching functions
generating the binary-state dynamics. The key underpinning of our method is
an expectation-maximization based algorithm to maximize the probability
(likelihood) that there is a link between an arbitrary pair of nodes in
the network. As a result, a feature that is particularly appealing from the
standpoint of network reconstruction arises: a distinct gap
between the probability values that correspond to actual links and
those associated with non-existent links. Statistical inference theory
also enables us to obtain an explicit formula for placing a threshold
in the gap so that the actual and non-existent links can be distinguished
unambiguously in an automated fashion. It is this feature that leads to
the superior performance of our statistical inference based methodology
as compared with those of the previous methods. In particular, we
demonstrate, using a large number of combinations of binary dynamical
processes and complex network topologies, that our method is capable
of reconstructing the network structure based solely on binary time series
with unprecedented accuracy, regardless of the nature of the intrinsic
switching functions generating the binary state dynamics. Additional
features of our methodology are effectively parameter free and robustness
against stochastic fluctuations in the data. While our method is
articulated for network structural reconstruction and hence does not
address the issue of identifying the underlying dynamical processes, it
represents a practically useful addition to the toolbox of reconstructing
complex networks structure and dynamics, which is being expanded at a
rapid pace by many research groups.

\acknowledgments

The authors thank Dr.~Z.-S. Shen for discussions. This work was supported
by NSFC under Grant Nos.~61473001 and 11331009, and partially supported by
the Young Talent Funding of Anhui Provincial Universities under Grant
No.~gxyqZD2017003. YCL would like to acknowledge support from the Vannevar
Bush Faculty Fellowship program sponsored by the Basic Research Office of
the Assistant Secretary of Defense for Research and Engineering and
funded by the Office of Naval Research through Grant No.~N00014-16-1-2828.

\section*{Appendix}

The basic structural parameters of the five empirical networks used in
the numerical demonstration are summarized in Tab.~\ref{table3}. The eight
binary-state dynamical modes are summarized below with parameters given
in Ref.~\cite{LSWGL:2017}.

\paragraph*{(1) Voter model.} The voter model assumes that a node randomly
chooses and then adopts one of its neighbors' state at each time step.
If $m$ neighbors among total $k$ neighbors are in an active state, the
probabilities of being active and inactive are $m/k$ and $(k-m)/k$,
respectively~\cite{SR:2005}. Since the voter model can cause the
nodal states to converge into a stable state, we randomly initialize the
states of all nodes after each 100 time steps.

\paragraph*{(2) Kirman model.} In this model, each node changes its
state from 0 to 1 with the probability $c_1 + dm$ and the probability
associated with the opposite change is $c_2 + d(k - m)$, where the
parameters $c_1$ and $c_2$ quantify the individual action that is
independent of the states of the neighbors and $d$ characterizes the
action of copying from neighbors' state~\cite{Kirman:1993}. In our
computations, we set $c_1=0.1$, $c_2=0.1$ and $d=0.08$.

\paragraph*{(3) Ising model.} This is the classic paradigm for understanding
ferromagnetism at the microscopic level of spins. Each node switches its
state from 0 to 1 with the probability $[1+e^{\beta (k-2m)/k}]^{-1}$ and from
state 1 to 0 with the probability $[e^{\beta(k-2m)/k}]/[1+e^{\beta(k-2m)/k}]$,
where $\beta=2$ characterizes the combining effect of temperature and
ferromagnetic interaction~\cite{KRB:book}.

\paragraph*{(4) SIS model.} This model describes the epidemic process of
disease spreading with infection and recovery. A susceptible individual
can be infected with probability $1-(1-\lambda)^m$ (from state 0 to 1) at
each time step, and an infected node can recover to the susceptible state
at the recovery rate $\mu$, where $\lambda$ is the transmission
rate~\cite{PCVV:2015}. In our simulations, we set $\lambda=0.5$ and
$\mu=0.5$ if the average degree is smaller than 10; otherwise we choose
$\lambda=0.35$ and $\mu=0.5$.

\begin{table}
\centering
\small
\caption{ {\bf Structural parameters of the five empirical networks used
in our numerical simulations}. The parameters $N$ and $E$ are the total
numbers of nodes and links, respectively, $C$ and $r$ are the clustering
and assortative coefficients, respectively, $H$ is the degree heterogeneity
defined as $H=\langle k^2\rangle/\langle k\rangle^2$.}
\begin{tabular}{ccccccc}
\hline
        Network &N  &E   &$\langle k\rangle$ &C &r &H         \\ \hline
        Karate &34  &78  & 4.5882 &0.5879 &-0.4756 &1.6933 \\
        Dolphins &62 &159 & 5.129  &0.2901 &-0.0718 &1.3255 \\
        Polbooks &105 &441 &8.40 &0.4875 &-0.1279 &1.4207\\
        Football &115 &613 &10.6609 &0.4032 &0.1624&1.0069\\
        Email &1133 &5451 &9.6222 &0.2540 &0.0782 &1.9421 \\ \hline
\end{tabular}\label{table3}
\end{table}

\paragraph*{(5) Game model.} For evolutionary game dynamics on complex
networks~\cite{SF:2007},  a player (a node) can be a cooperator (active -
the 1 state) or a defector (inactive - the 0 state). A player plays with
each of his/her neighbors using one chosen strategy at every time step.
The players obtain payoff $a$ ($d$) if both choose to cooperate (defect).
If one player cooperates while the other defects, the cooperator will
obtain low payoff $b$, while the defector will gain higher
payoff $c$. The payoff of a player is the sum of payoffs from playing
game with all its neighbors. A player switches the strategy with a
probability that depends on the payoff it may gain in the next round under
the current circumstance. Each player switches its state from 0 to 1 with
the probability $[\alpha+e^{\frac{\beta}{k}([(a-c)(k-m)+(b-d)m])}]^{-1}$
and from state 1 to 0 with the probability
$[\alpha+e^{\frac{\beta}{k}([(c-a)(k-m)+(d-b)m])}]^{-1}$, where $\alpha$
qualifies the willingness for a player to change its strategy according
to those of its neighbors, and $\beta$ is associated with the effect of
the expected payoff. We choose $a=b=5$, $c=d=0$, $\alpha=0.1$, and $\beta=1$
in our simulations.

\paragraph*{(6) Language model.} In this model, the two states denote two
different language choices of a person. The transition probability from
the primary to the secondary language is proportional to the fraction of
speakers in the neighbors with the power $\alpha$, multiplied by the parameter
$s$ (or $1-s$) according to the respective language~\cite{AS:2003}, where
$\alpha=0.7$ and $s=0.5$. Because of the problem of converging to a stable
state, we randomly initialize the states of all nodes after every 100 time
steps.

\paragraph*{(7) Threshold model.}
This is a deterministic model, where a node becomes active if the fraction
of active neighbors $m/k$ is larger than the threshold $1/2$, and no
recovery transformation is permitted~\cite{Granovetter:1978}. Due to the
problem of fast convergence to a stable state from any initial condition,
we randomly initialize the states of all nodes after every 5 time steps.

\paragraph*{(8) Majority-voter model.}
In this model, a node tends to align with the majority state of its
neighbors, with $Q$ being the probability of misalignment~\cite{Oliveira:1992}.
We set $Q=0.3$ and randomly initialize the states of all nodes after
every 10 time steps to overcome the difficulty of fast convergence to a
stable state.


%
\end{document}